\documentclass[12pt]{iopart}

\usepackage{cite}
\usepackage{graphicx}
\usepackage{units}
\usepackage[colorlinks,
breaklinks,
citecolor=blue,
linkcolor=blue,
]{hyperref}
\begin{document}

\title[Dynamical $N$-photon bundle emission]{Dynamical $N$-photon bundle emission}

\author{F. Zou\textsuperscript{1,2,3}, Y. Li\textsuperscript{4,2} and J.-Q. Liao\textsuperscript{1,2,*}}

\address{\textsuperscript{1} Key Laboratory of Low-Dimensional Quantum Structures and Quantum Control of
Ministry of Education, Key Laboratory for Matter Microstructure and Function of Hunan Province, Department of Physics, Hunan Normal University, Changsha 410081, China}
\address{\textsuperscript{2} Synergetic Innovation Center for Quantum Effects and Applications, Hunan Normal University, Changsha 410081, China}
\address{\textsuperscript{3} Beijing Computational Science Research Center, Beijing 100193, China}
\address{\textsuperscript{4} Center for Theoretical Physics and School of Science, Hainan University, Haikou 570228, China}
\address{\textsuperscript{*} Author to whom any correspondence should be addressed.}
\ead{jqliao@hunnu.edu.cn}

\vspace{10pt}

\begin{abstract}
Engineering multiphoton resources is of importance in quantum metrology, quantum lithography, and biological sensing. Here we propose a concept of dynamical emission of $N$ strongly-correlated photons. This is realized in a circuit quantum electrodynamical system driven by two Gaussian-pulse sequences. The underlying physical mechanism relies on the stimulated Raman adiabatic passage that allows efficient and selective preparation of target multiphoton states. Assisted by the photon decay, a highly pure $N$-photon bundle emission takes place in this system. In particular, the dynamical $N$-photon bundle emission can be tuned by controlling the time interval between consecutive pulses so that the device behaves as an $N$-photon gun, which can be triggered on demand. Our work opens up a route to achieve multiphoton source devices, which have wide potential applications in quantum information processing and quantum metrology.
\end{abstract}

\vspace{2pc}

\submitto{\NJP}

\maketitle
%\ioptwocol

\section{Introduction}
Multi-photon sources~\cite{satzinger2018Quantum,chu2018Creation,Gokhale2020Epitaxial} have wide applications in quantum communication~\cite{kimble2008Quantum},  lithography\cite{dangelo2001TwoPhoton}, spectroscopy~\cite{lopezcarreno2015Exciting,dorfman2016Nonlinear}, quantum metrology~\cite{giovannetti2006Quantum}, and quantum biology~\cite{denk1990Twophoton,horton2013Vivo}. Multi-photon bundle emission~\cite{munoz2014Emitters,strekalov2014Bundle,munoz2018Filtering,Jiang2022Multiple,bin2021ParitySymmetryProtected}, as one of the physical mechanisms for generation of multi-photon sources, has attracted considerable interest in the past few years. In general, an advanced preparation of the $N$-photon states is necessary for the implementation of $N$-photon bundle emission. To date, many methods have been proposed to generate $N$-photon states in various physical systems, such as cavity quantum electrodynamical (QED) systems~\cite{law1996Arbitrary,Huang2014Photon,munoz2014Emitters,strekalov2014Bundle,
munoz2018Filtering,Jiang2022Multiple,bin2021ParitySymmetryProtected,Cosacchi2021photon,Camacho2021Multiphoton}, circuit-QED systems~\cite{ma2021Antibunched}, coupled photon-atom systems~\cite{dousse2010Ultrabright,Yasutomo2011Spontaneous,koshino2013Implementation,callsen2013Steering,muller2014Ondemand,
munoz2015Enhanced,chang2016Deterministic,hargart2016Cavityenhanced,sanchez-burillo2016Full}, Rydberg atomic ensembles~\cite{bienias2014Scattering,maghrebi2015Coulomb}, waveguide systems~\cite{gonzalez-tudela2015Deterministic,douglas2016Photon,gonzalez-tudela2017Efficient}, Kerr cavity systems~\cite{Liao2010Correlated}, and cavity optomechanical systems~\cite{Liao2013Correlated,qin2019Emission}. Particularly appealing, $N$-photon bundle emission based on the Mollow physics of resonance fluorescence~\cite{munoz2014Emitters,strekalov2014Bundle,munoz2018Filtering,Jiang2022Multiple} and parity symmetry of quantum Rabi model~\cite{bin2021ParitySymmetryProtected} have recently been proposed. Meanwhile, the $N$-photon bundle emission has been realized experimentally in a dc-biased superconducting circuit~\cite{Menard2022Emission}. We also note that many schemes have been proposed to realize $N$-phonon bundle emission in various physical systems, such as a strongly driven nitrogen-vacancy center coupled to a mechanical resonator~\cite{dong2019Multiphonon}, an acoustic cavity quantum electrodynamical (QED) system~\cite{bin2020Phonon}, and a generalized quantum Rabi system~\cite{deng2021Motional}. The generation of $N$-phonon states is of paramount importance for quantum communication~\cite{Gustafsson2014Propagating,Kuzyk2018Scaling}, acoustic quantum precision measurement~\cite{Toyoda2015Hong,Zhang2018NOON}, and ultrasensitive detection~\cite{Chu2011Ultrasensitive}.

Physically, to realize an efficient $N$-photon emission, it is desired to generate a perfect $N$-photon number states in advance. From an application perspective, it is expected to create the states in a dynamical and deterministic manner. Here, the dynamical requirement confirms that the photon states can be generated under control and that the preparation of the photon states can be completed on demand. Meanwhile, the deterministic way ensures the high efficiency of the state generation.

Here, we propose a concept of dynamical $N$-photon bundle emission and  present a feasible scheme to generate dynamical $N$-photon ($N=2,3$) bundle emission in a circuit-QED system, in which a microwave resonator is longitudinally coupled to a qubit driven by two Gaussian-pulse sequences. Based on the feature of stimulated Raman adiabatic passage (STIRAP)~\cite{gaubatz1990Population,bergmann1998Coherent,vitanov2017Stimulated}, the population transfer between zero- and $N$-photon states can be realized. Combined with photon decay of the resonator, the $N$-photon states can be emitted as photon bundles. This means that the dynamical $N$-photon bundle emission is realized via the dissipative channel. This dynamical $N$-photon bundle emission can be tuned on demand by controlling the time interval between consecutive pulses, then the device behaves as an $N$-photon gun, with wide applications for quantum science and technology.

\section{Model}
\begin{figure}[t]
\center
\includegraphics[width=0.6 \textwidth]{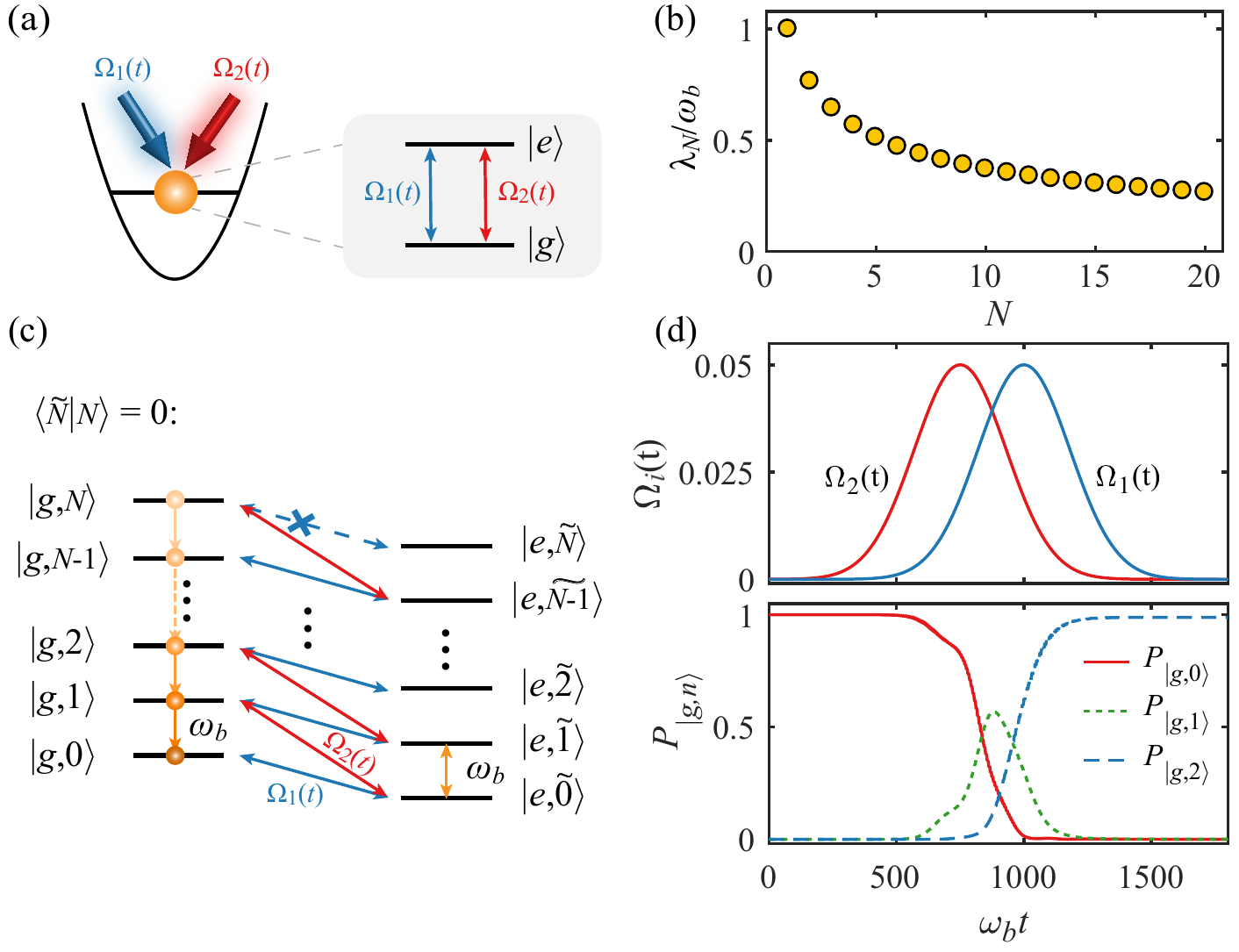}
\caption{(Color online) (a) Schematic of the circuit-QED system, which is composed of a qubit longitudinally coupled to a microwave resonator. The qubit is driven by two consecutive driving pulses $\Omega_{1}(t)$ and $\Omega_{2}(t)$. (b) The minimal positive value of $\lambda_{N}/\omega_{b}$ satisfying Eq.~(\ref{miniequ}) at different $N$. (c) The resonant transition chain of the approximate Hamiltonian $H_{\mathrm{app}}^{[N]}$ for $\lambda=\lambda_{N}$, without transition between states $\vert g,N\rangle$ and $\vert e,\tilde{N}\rangle$. (d) Two driving pulses $\Omega_{i=1,2}(t)$ ($m=0$) and the state populations $P_{\vert g,n\rangle}(t)=\vert\langle g,n\vert\psi(t)\rangle\vert^{2}$ as functions of the time $\omega_{b}t$. Other parameters are $\Omega_{0}/\omega_{b}=0.05$, $\omega_{b}\sigma=180$, $\omega_{b}t_{1}=1000$, $\omega_{b}t_{2}=750$, $\lambda/\omega_{b}\approx0.765$, $\Delta_{1}=-\lambda^{2}/\omega_{b}$, and $\Delta_{2}=-\lambda^{2}/\omega_{b}-\omega_{b}$.}
\label{Fig1}
\end{figure}

We consider a circuit-QED system, which is composed of a qubit longitudinally coupled to a microwave resonator [Fig.~\ref{Fig1}(a)]. The Hamiltonian of the system reads ($\hbar=1$)
\begin{equation}\label{Hs}
H_{s}=\omega_{b}b^{\dagger}b+\omega_{0}\sigma_{+}\sigma_{-}+\lambda\sigma_{+}\sigma_{-}(b^{\dagger}+b),
\end{equation}
where $b^{\dagger}$ ($b$) is the creation (annihilation) operator of the microwave resonator with resonant frequency $\omega_{b}$. The operators $\sigma_{+}=\vert e\rangle\langle g\vert$ and $\sigma_{-}=\vert g\rangle\langle e\vert$ are, respectively, the raising and lowering operators of the qubit, with transition frequency $\omega_{0}$ between the excited state $\vert e\rangle$ and the ground state $\vert g\rangle$. The last term in Eq.~(\ref{Hs}) denotes the interaction between the qubit and the resonator with the coupling strength $\lambda$. Note that in Eq.~(\ref{Hs}) we have displaced the photonic field by introducing a driving $\lambda(b^{\dagger}+b)/2$ to the resonator, as an assistance of the longitude qubit-resonator coupling $\lambda\sigma_{z}(b^{\dagger}+b)/2$, such that the generated photon bundles are not displaced by the qubit. This point can be seen based on the relation $\lambda\sigma_{z}(b^{\dagger}+b)/2+\lambda(b^{\dagger}+b)/2=\lambda\sigma_{+}\sigma_{-}(b^{\dagger}+b)$. Note that the physical effect induced by the longitudinal qubit-resonator interaction has been studied both theoretically~\cite{Romero2012Ultrafast,Didier2015Fast,Billangeon2015Circuit,Billangeon2015Scalable,Richer2016Circuit,Richer2017Inductively} and experimentally~\cite{Touzard2019Gated,Ikonen2019Qubit,Hacohen2016Quantum,Eddins2018Stroboscopic}. In the case of without the driving $\lambda(b^{\dagger}+b)/2$ of the resonator, the eigenstates of the Hamiltonian $H_{s}^{\prime}=\omega_{b}b^{\dagger}b+\omega_{0}\sigma_{+}\sigma_{-}+\lambda\sigma_{z}(b^{\dagger}+b)/2$ are $\vert e\rangle D(-\lambda/2\omega_{b})\vert n\rangle$ and $\vert g\rangle D(\lambda/2\omega_{b})\vert n\rangle$, respectively. Here $D(\pm\lambda/2\omega_{b})\vert n\rangle$ is the displaced-photon-number state, with $D(\pm\lambda/2\omega_{b})=\exp[\pm(\lambda/2\omega_{b})(b^{\dagger}-b)]$ being the displacement operator. In this case, the generated photon bundles will be coherently displaced.
%%%%%%%%%%%%%%%%%%%%%
%%%%%%%%%%%%%%%%%%%%%%%

To study the dynamical bundle emission of $n$ strongly-correlated photons in this system, the qubit is driven by two Gaussian-pulse sequences with corresponding carrier frequencies $\omega_{1}$ and $\omega_{2}$. Each Gaussian wave packet of the two Gaussian-pulse sequences has the same amplitude and width. The driving Hamiltonian is
\begin{equation}
H_{d}=\Omega_{1}(t)\sigma_{+}e^{-i\omega_{1}t}+\Omega_{2}(t)\sigma_{+}e^{-i\omega_{2}t}+\mathrm{H.c.},
\end{equation}
with
\begin{equation}
\Omega_{i}(t)=\Omega_{0}\sum_{m=0}^{N_{0}}\exp[-(t-t_{i}-mT)^{2}/(2\sigma^{2})],\quad i=1,2.
\end{equation}
Here, $\Omega_{0}$ and $\sqrt{2}\sigma$ are the amplitude and width of each Gaussian wave packet, respectively. The $t_{i}+mT$ denotes the time when the pulse $\Omega_{i}(t)$ reaches its maximum value, where $m$ is an integer and $T$ is the time interval between consecutive pulses. $N_{0}+1$ is the number of the pulses. In the rotating frame with respect to $\omega_{0}\sigma_{+}\sigma_{-}$, the total Hamiltonian of the system becomes
\begin{equation}\label{H}
H=\omega_{b}b^{\dagger}b+\lambda\sigma_{+}\sigma_{-}(b^{\dagger}+b)+[\Omega_{1}(t)\sigma_{+}e^{-i\Delta_{1}t}+\Omega_{2}(t)\sigma_{+}e^{-i\Delta_{2}t}+\mathrm{H.c.}],
\end{equation}
where $\Delta_{i}=\omega_{i}-\omega_{0}$ ($i=1,2$) is the detuning of the driving carrier frequency $\omega_{i}$ with respect to the transition frequency $\omega_{0}$ of the qubit.

For the Hamiltonian $H_{0}=\omega_{b}b^{\dagger}b+\lambda\sigma_{+}\sigma_{-}(b^{\dagger}+b)$, its eigenvalues are $E_{g,n}=n\omega_{b}$ and $E_{e,n}=n\omega_{b}-\lambda^{2}/\omega_{b}$, with the corresponding eigenstates $\vert g,n\rangle$ and $\vert e,\tilde{n}\rangle$. Here, $\vert n\rangle$ ($n=0,1,2,\cdots$) is the photon number state and $\vert\tilde{n}\rangle\equiv D(-\lambda/\omega_{b})\vert n\rangle$ is the displaced-photon-number state, with $D(-\lambda/\omega_{b})=\exp[-(\lambda/\omega_{b})(b^{\dagger}-b)]$ being the displacement operator. The Hamiltonian~(\ref{H}) can be expressed using the eigenstates of $H_{0}$ as
\begin{eqnarray}
H &=& \sum_{n}(E_{g,n}\vert g,n\rangle\langle g,n\vert+E_{e,n}\vert e,\tilde{n}\rangle\langle e,\tilde{n}\vert) \nonumber\\
&&+[\sum_{i=1,2}\sum_{n,m}\Omega_{i}(t)e^{-i\Delta_{i}t}\langle\tilde{n}\vert m\rangle\vert e,\tilde{n}\rangle\langle g,m\vert+\mathrm{H.c.}].
\end{eqnarray}
Here the Franck-Condon factor $\langle\tilde{n}\vert m\rangle=\langle n\vert D(\lambda/\omega_{b})\vert m\rangle$~\cite{deoliveira1990Properties} can be calculated by
\begin{equation}
\langle\tilde{n}\vert m\rangle=\cases{\sqrt{\frac{n!}{m!}}e^{-\frac{\beta^{2}}{2}}(-\beta)^{m-n}L_{n}^{m-n}\left(\beta^{2}\right), &$n\le m$,\\
\sqrt{\frac{m!}{n!}}e^{-\frac{\beta^{2}}{2}}\beta^{n-m}L_{m}^{n-m}\left(\beta^{2}\right), &$n>m$,\\}
\end{equation}
where $\beta=\lambda/\omega_{b}$ and $L_{n}^{m}(x)$ are the associated Laguerre polynomials. In the rotating frame with respect to $H_{0}$, the Hamiltonian becomes
\begin{equation}
H_{I}=\sum_{i=1,2}\sum_{n,m}\Omega_{i}(t)e^{i(\delta E_{n,m}-\Delta_{i})t}\langle\tilde{n}\vert m\rangle\vert e,\tilde{n}\rangle\langle g,m\vert+\textrm{H.c.},
\end{equation}
where we introduce the parameter
\begin{equation}
\delta E_{n,m}=E_{e,n}-E_{g,m}=(n-m)\omega_{b}-\lambda^{2}/\omega_{b}.
\end{equation}

\section{Approximate Hamiltonian}
In the resolved-sideband regime (i.e., the photon frequency $\omega_{b}$ is much larger than the decay rate $\gamma$ of the qubit, $\omega_{b}/\gamma\gg1$), we choose the proper driving carrier frequencies $\omega_{1}$ and $\omega_{2}$ such that the detunings $\Delta_{1}$ and $\Delta_{2}$ satisfy the resonant conditions:
\begin{eqnarray}
\Delta_{1}&=&\delta E_{n,n}=E_{e,n}-E_{g,n}=-\lambda^{2}/\omega_{b}, \nonumber\\
\Delta_{2}&=&\delta E_{n,n+1}=E_{e,n}-E_{g,n+1}=-\lambda^{2}/\omega_{b}-\omega_{b}.
\end{eqnarray}
In this case, the Hamiltonian $H_{I}$ can be decomposed into two parts: $H_{I}=H^{\prime}_{I}+H^{\prime\prime}_{I}$, where $H^{\prime}_{I}$ corresponds to the resonant-transition part
\begin{equation}
H^{\prime}_{I}=\sum_{n}[\Omega_{1}(t)\langle\tilde{n}\vert n\rangle\vert e,\tilde{n}\rangle\langle g,n\vert
+\Omega_{2}(t)\langle\tilde{n}\vert n+1\rangle\vert e,\tilde{n}\rangle\langle g,n+1\vert+\mathrm{H.c.}],
\end{equation}
and $H^{\prime\prime}_{I}$ denotes the off-resonant-transition part
\begin{equation}
H^{\prime\prime}_{I}=\sum_{i=1,2}\sum_{n,m}\!^{\prime}[\Omega_{i}(t)e^{i\delta_{i}(n,m)t}\langle\tilde{n}\vert m\rangle\vert e,\tilde{n}\rangle\langle g,m\vert+\textrm{H.c.}].
\end{equation}
Here the primed summation in $H^{\prime\prime}_{I}$ eliminates these terms of $H^{\prime}_{I}$, and the off-resonance detunings are $\delta_{i}(n,m)=\delta E_{n,m}-\Delta_{i}$ for $i=1,2$. When the off-resonance detunings $\vert\delta_{i}(n,m)\vert$ ($i=1,2$) are much greater than $\Omega_{0}\vert\langle\tilde{n}\vert m\rangle\vert$, then the high-frequency oscillating Hamiltonian~$H^{\prime\prime}_{I}$ can be neglected by the rotating-wave approximation, namely, $H_{I}\approx H^{\prime}_{I}$. Since $\vert\langle\tilde{n}\vert m\rangle\vert\leq1$~\cite{Xu2013Dark}, the condition of neglecting $H^{\prime\prime}_{I}$ is reduced to $\Omega_{0}\ll\omega_{b}$. In addition, we choose the proper coupling strength $\lambda=\lambda_{N}$, where $\lambda_{N}$ is the minimal positive value for satisfying the equation~\cite{Xu2013Dark}
\begin{equation}\label{miniequ}
\langle\tilde{N}\vert N\rangle=\exp\left[-\lambda_{N}^{2}/\left(2\omega_{b}^{2}\right)\right]L_{N}^{0}\left(\lambda_{N}^{2}/\omega_{b}^{2}\right)=0,
\end{equation}
with $N$ being a positive integer. Equation~(\ref{miniequ}) denotes that the transition matrix element $\Omega_{1}(t)\langle\tilde{N}\vert N\rangle$ between the states $\vert g,N\rangle$ and $\vert e,\tilde{N}\rangle$ is zero. Hence, the dimension of the photon Hilbert space can be truncated up to $n=N$. To investigate how the coupling strength $\lambda_{N}$ depends on $N$, we plot in Fig.~\ref{Fig1}(b) the coupling strength $\lambda_{N}/\omega_{b}$ as a function of the index $N$. It can be seen that the coupling strength $\lambda_{N}/\omega_{b}$ decreases as the index $N$ increases.

Under the above mentioned conditions, the system can be well described by the approximate Hamiltonian
\begin{equation}
H_{\mathrm{app}}^{[N]}=\sum_{n=0}^{N-1}[\Omega_{1}(t)\langle\tilde{n}\vert n\rangle\vert e,\tilde{n}\rangle\langle g,n\vert
+\Omega_{2}(t)\langle\tilde{n}\vert n+1\rangle\vert e,\tilde{n}\rangle\langle g,n+1\vert+\textrm{H.c.}], \label{Heff}
\end{equation}%
which describes the resonant transition chain: $\vert g,0\rangle\leftrightarrow\vert e,\tilde{0}\rangle\leftrightarrow\vert g,1\rangle\leftrightarrow\vert e,\tilde{1}\rangle\leftrightarrow\cdots\vert e,\widetilde{N-1}\rangle\leftrightarrow\vert g,N\rangle$, as shown in Fig.~\ref{Fig1}(c). In Fig.~\ref{Fig1}(d), we show the two pulsed driving fields $\Omega_{i=1,2}(t)$ ($m=0$) and the state populations $P_{\vert g,n\rangle}(t)=\vert\langle g,n\vert\psi(t)\rangle\vert^{2}$ as functions of $\omega_{b}t$ at $\lambda/\omega_{b}=\lambda_{2}/\omega_{b}\approx0.765$, i.e., corresponding to $\langle\tilde{2}\vert 2\rangle=0$. Here we consider that the initial state of the system is $\vert g,0\rangle$. We see a perfect population transfer from $\vert g,0\rangle$ to $\vert g,2\rangle$ in the absence of the system dissipation when the two Gaussian pulses $\Omega_{i=1,2}(t)$ satisfies an adiabatic evolution. This state transfer is determined by the physical mechanism of STIRAP~\cite{gaubatz1990Population,bergmann1998Coherent,vitanov2017Stimulated}.
%%%%%%%%%%%%%%%%%%%%%
\begin{figure}[t]
\center
\includegraphics[width=0.6 \textwidth]{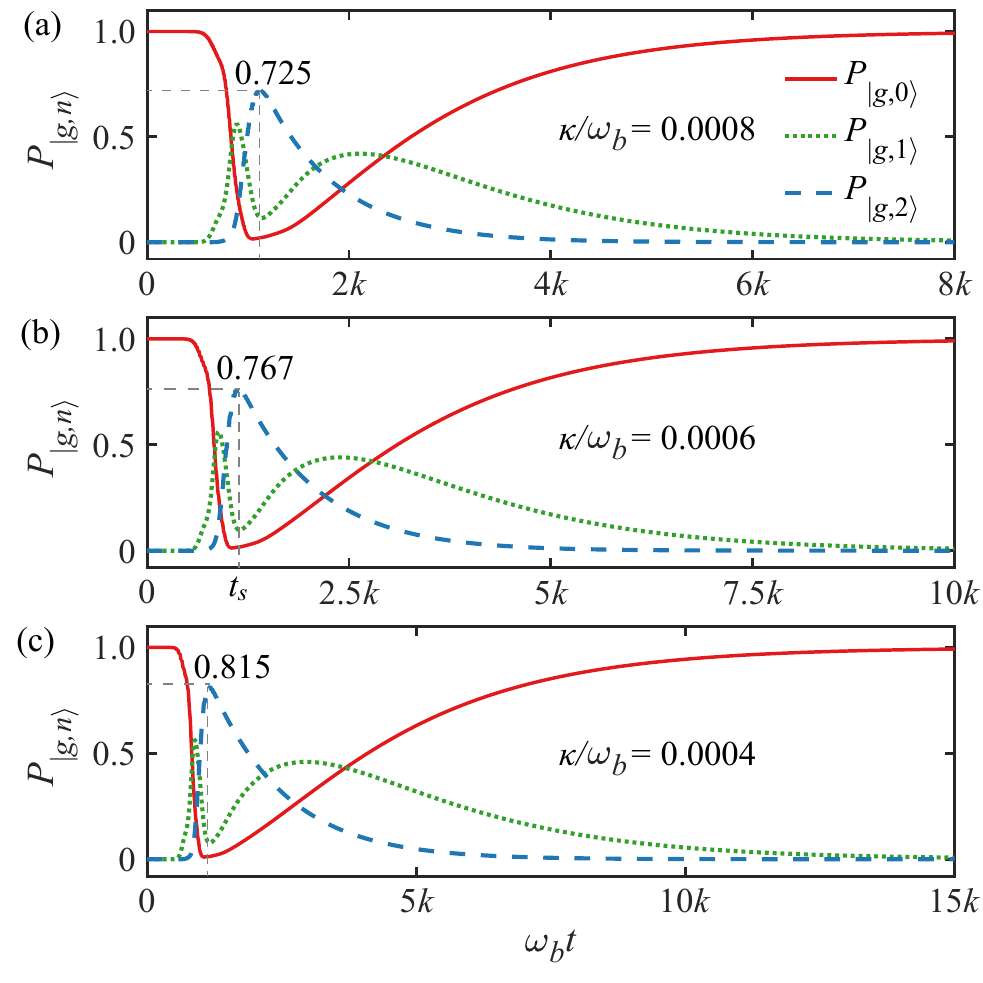}
\caption{(Color online) The state populations $P_{\vert g,n\rangle}(t)$ as functions of $\omega_{b}t$ when the decay rate of the resonator is taken as (a) $\kappa/\omega_{b}=0.0008$, (b) $\kappa/\omega_{b}=0.0006$, and (c) $\kappa/\omega_{b}=0.0004$. The horizontal axis is scaled by a factor of $k=1000$. Here we take $\gamma/\omega_{b}=0.002$, and other parameters are the same as those given in Fig.~\ref{Fig1}.}
\label{Fig2}
\end{figure}
%%%%%%%%%%%%%%%%%%%%%%%

To include the dissipation of the system, we consider the case where the resonator and the qubit are connected with two individual vacuum baths. Then the system is governed by the dressed-state master equation working in the ultrastrong-coupling regime ~\cite{Settineri2018Dissipation,Diaz2019Ultrastrong,Kockum2019Ultrastrong,Boite2020Theoretical,Ridolfo2012Photon,scully1997Quantum,Hu2015Quantum} (see Appendix),
\begin{equation}
\frac{d\rho(t)}{dt}=i[\rho(t),H]+\kappa\mathcal{L}[b+(\lambda/\omega_{b})\sigma_{+}\sigma_{-}]\rho(t)+\gamma\mathcal{L}[\sigma_{-}]\rho(t),\label{SME}
\end{equation}
where $H$ is given by Eq.~(\ref{H}) and $\kappa$ ($\gamma$) is the decay rate of the resonator (qubit). The Lindblad superoperators are defined by
\begin{equation}
\mathcal{L}[o]\rho(t)=[2o\rho(t)o^{\dagger}-o^{\dagger}o\rho(t)-\rho(t) o^{\dagger}o]/2,
\end{equation}
with $o=b+(\lambda/\omega_{b})\sigma_{+}\sigma_{-}$ and $\sigma_{-}$. In Fig.~\ref{Fig2}, we show the state populations $P_{\vert g,n\rangle}(t)$ as functions of $\omega_{b}t$ at various values of $\kappa/\omega_{b}$. Here we only show one period of the STIRAP pulses. It can be seen from Figs.~\ref{Fig2}(a)-\ref{Fig2}(c) that the maximum value of the population $P_{\vert g,2\rangle}(t)$ is less than 1 in the presence of dissipation. Owing to the photon dissipation, the two photons are emitted out of the cavity, the system is then brought back to the initial state $\vert g,0\rangle$. The state $\vert g,2\rangle$ is again generated for the next Gaussian pulse. Hence, the two Gaussian-pulse sequences lead to a transition $\vert g,0\rangle\leftrightarrow\vert g,2\rangle$ under the assistant of the photon dissipation. In addition, we find that the maximum value of the population $P_{\vert g,2\rangle}(t)$ decreases as $\kappa/\omega_{b}$ increases. Meanwhile, for a smaller photon dissipation rate, a longer relaxation time is needed to reach the steady state.

It should be noted that, in principle, the lowing operator associated with the coupled system in the ultrastrong-coupling regime should be $b+(\lambda/\omega_{b})\sigma_{+}\sigma_{-}$ rather than the annihilation operator $b$ of the resonator. However, in the STIRAP scheme, the system is mainly in the state $\vert g,n\rangle$, and the population of the state $\vert e,\tilde{n}\rangle$ is approximately 0. For example, we can see from Fig.~\ref{Fig1}(d) that the population is $P_{\vert g,0\rangle}+P_{\vert g,1\rangle}+P_{\vert g,2\rangle}\approx1$ during the STIRAP, which indicates that $P_{\vert e,\tilde{n}\rangle}\approx0$, i.e., $\langle\sigma_{+}\sigma_{-}\rangle\approx0$. In this case, it is reasonable to approximately calculate the photon statistics with the annihilation operator $b$ instead of the displaced operator $b+(\lambda/\omega_{b})\sigma_{+}\sigma_{-}$.

\section{Dynamical $N$-photon bundle emission}

%%%%%%%%%%%%%%%%%%%%%
\begin{figure}[t]
\center
\includegraphics[width=0.6 \textwidth]{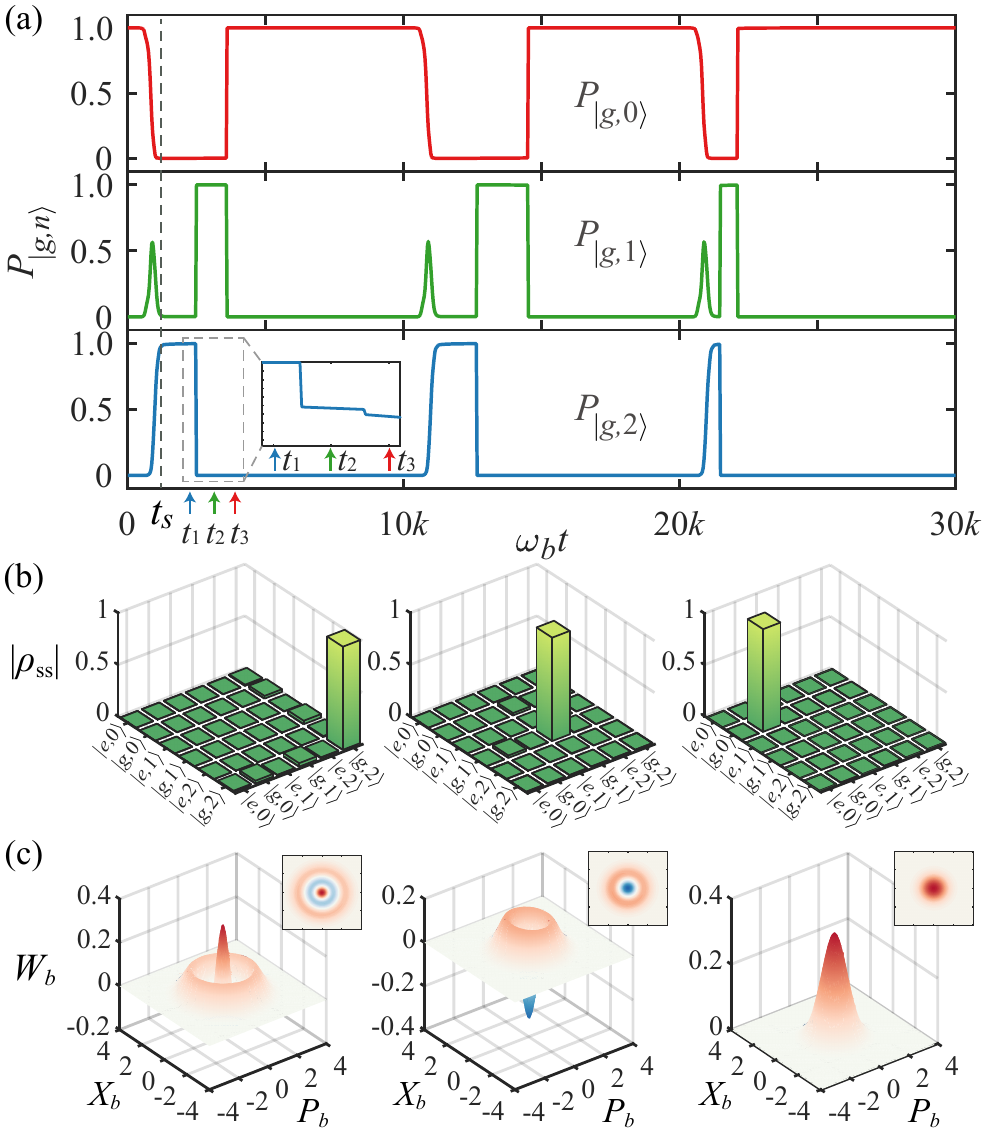}
\caption{(Color online) (a) Quantum trajectory of the state populations $P_{\vert g,n\rangle}(t)$, showing the dynamical two-photon bundle emission. (b) The full density matrix elements $|\rho_{\mathrm{ss}}|$ of the system at three different moments with the color arrows in the trajectory, showing the cascade-photon-emission process $|g,2\rangle\rightarrow|g,1\rangle\rightarrow|g,0\rangle$. (c) The Wigner functions $W_{b}$ of the reduced density matrix of the photon mode at three different moments. Here we take $\kappa/\omega_{b}=0.0006$, $\gamma/\omega_{b}=0.002$, $\omega_{b}T=10000$, and other parameters are given in Fig.~\ref{Fig1}.}
\label{Fig3}
\end{figure}
%%%%%%%%%%%%%%%%%%%%%%%

To confirm the dynamical $N$-photon bundle emission, we employ a quantum Monte-Carlo approach to follow individual trajectory of the system~\cite{munoz2014Emitters,bin2020Phonon,bin2021ParitySymmetryProtected}. Figure~\ref{Fig3}(a) shows a quantum trajectory of dynamical two-photon bundle emission under the driving of two Gaussian-pulse sequences. Here $P_{\vert g,n\rangle}(t)$ are the populations of the states $\vert g,n\rangle$ for $n=0,1,2$. Initially, we consider that the system is in state $\vert g,0\rangle$. At time $t\approx t_{s}$, the system is transferred via the STIRAP from the initial state $\vert g,0\rangle$ to the two-photon state $\vert g,2\rangle$ with a prefect probability. The dissipation of the resonator triggers a quantum collapse of the system, from the two-photon state $\vert g,2\rangle$ to the one-photon state $\vert g,1\rangle$ with an almost unit probability, which causes the emission of the first photon. Subsequently, the system transits from the one-photon state $\vert g,1\rangle$ to the zero-photon state $\vert g,0\rangle$, completing the two-photon emission within the resonator lifetime. After the two-photon emission, the system goes back to the initial state $\vert g,0\rangle$. Next, the system is again prepared in $\vert g,2\rangle$ by the next Gaussian pulses, as the starting of the next emission of two photons. Hence, the dynamical two-photon bundle emission can be realized based on the sequential STIRAP. Here we choose the time $T$ between coterminous pulses is much greater than the lifetime $1/\kappa$ of the resonator such that the system can go back to the initial state $\vert g,0\rangle$ before the arrival of the next Gaussian pulses. In principle, the time duration between two STIRAPs can be controlled such that the emission of two photons can be triggered on demand.

We also prove the two-photon emission by checking both the density matrix and the Wigner function of the photon state. Figures~\ref{Fig3}(b) and~\ref{Fig3}(c) show the full density matrix elements $|\rho_{\mathrm{ss}}|$ of the system and the Wigner functions $W_{b}$ of the reduced density matrix of the photon mode at three different moments. It can be seen that the cascade-photon-emission process $|g,2\rangle\rightarrow|g,1\rangle\rightarrow|g,0\rangle$ occurs in a very short time window.

%%%%%%%%%%%%%%%%%%%%%
\begin{figure}[t]
\center
\includegraphics[width=0.6 \textwidth]{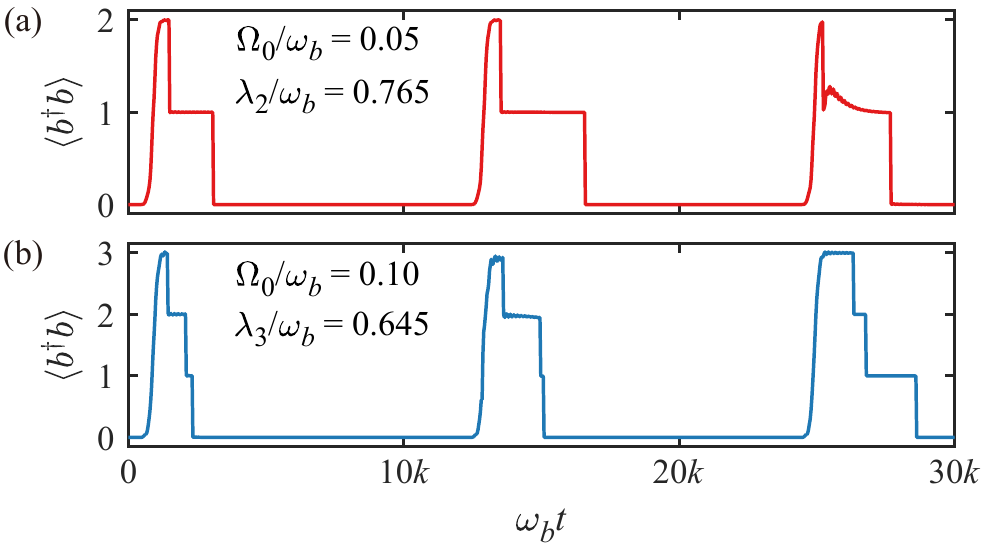}
\caption{(Color online) Quantum trajectory of the average photon number $\langle b^{\dagger}b\rangle(t)$ at (a) $\lambda/\omega_{b}\approx0.765$ and $\Omega_{0}/\omega_{b}=0.05$, showing the dynamical two-photon bundle emission; (b) $\lambda/\omega_{b}\approx0.645$ and $\Omega_{0}/\omega_{b}=0.1$, showing the dynamical three-photon bundle emission. Here we take $\omega_{b}T=12000$ and other parameters are given in Fig.~\ref{Fig2}.}
\label{Fig4}
\end{figure}
%%%%%%%%%%%%%%%%%%%%%%%
Figure~\ref{Fig4} shows a quantum trajectory of the average photon number $\langle b^{\dagger}b\rangle(t)$ at different values of $\lambda/\omega_{b}$ and $\Omega_{0}/\omega_{b}$. In Figs.~\ref{Fig4}(a) and~\ref{Fig4}(b), we observe that the photon mode evolves from $\vert N\rangle$ to $\vert 0\rangle$ ($N=2,3$) during each cycle of bundle emission. Owing to the dissipation of the resonator, the system undergoes a rapid cascade emission through the series of the Fock states $\vert m\rangle$ where $0\leq m\leq N$ in very short temporal windows. These results indicate that the dynamical $N$-photon ($N=2,3$) bundle emission is realized in this system under the combination of the sequential STIRAP and the photon dissipation.

We point out that during the $N$-photon bundle emission process, there is no wave-packet driving. Therefore, the bundle emission of the $N$ photons generated by the STIRAP scheme can be approximately understood as the $N$-photon emission physical process in a free cavity. Namely, for each bundle emission, there are two steps. The first step is the generation of $N$-photon Fock state in the cavity by the STIRAP, and the second step is the emitting of $N$ photons by the decay process of the free cavity. According to the quantum optics theory, we know that the emissions of the $N$ photons are independent processes and the $N$ emitted photons in the outside fields are in $N$ independent wave packets in frequency space [Its wave function in position space can be expressed as $\prod_{j=1}^{N}\varphi_{j}(x_{j})$ with $\varphi_{j}(x_{j})=\exp{\{-i\kappa[(x_{j}/v_{g})-t]/2\}}\theta[t-(x_{j}/v_{g})]$, where $\kappa$ is the decay rate of the free cavity, $x_{j}$ is the coordinate of the $j$th photon in the outside fields, $v_{g}$ is the group velocity of the photon in the outside fields, and $\theta(t)$ is the Heaviside step function]. This result is understandable in physics because the cavity is free. Therefore, we can define a single temporal mode (namely $N$ independent wave packets in frequency space) in the output fields to describe the state of the $N$ emitted photons.

%%%%%%%%%%%%%%%%%%%%%
\begin{figure}[t]
\center
\includegraphics[width=0.6 \textwidth]{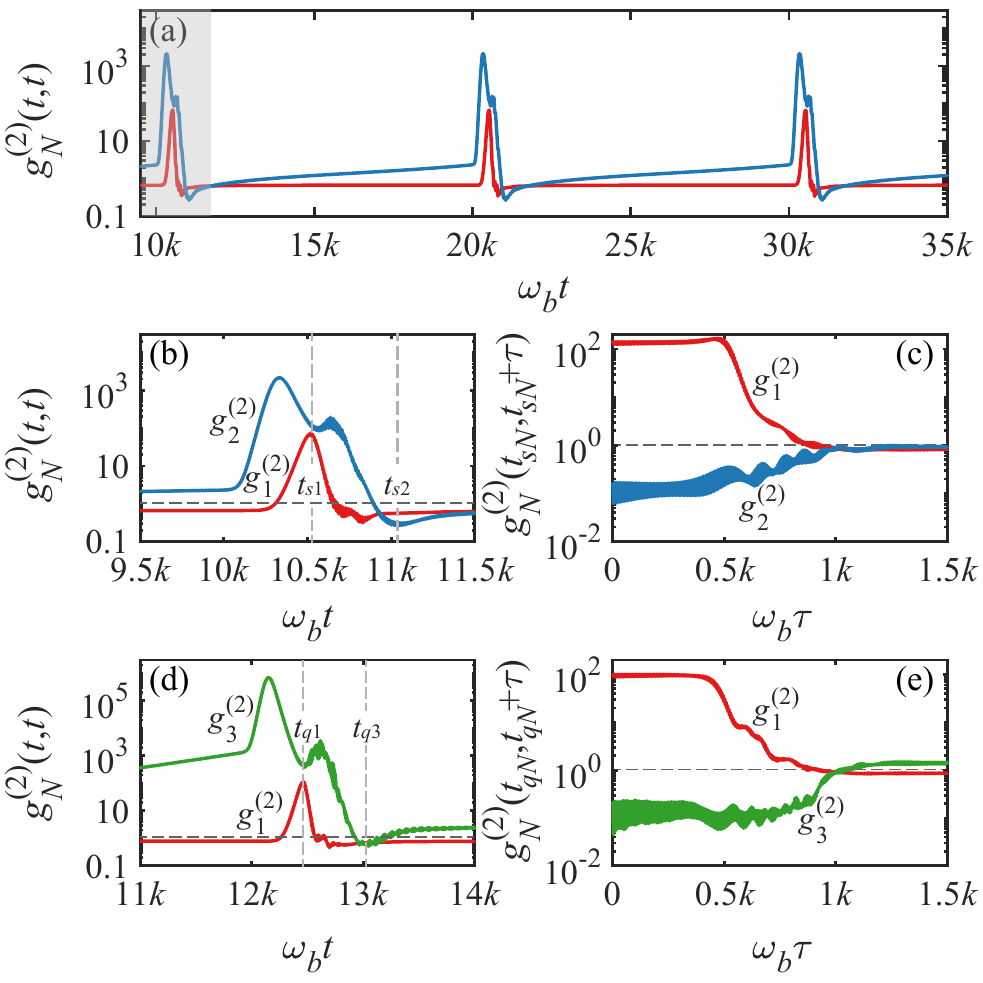}
\caption{(Color online) (a) The equal-time second-order correlation functions $g_{N}^{(2)}(t,t)$ as functions of $\omega_{b}t$ with $N=1$ (red) and $N=2$ (blue) when $\lambda/\omega_{b}\approx0.765$, $\Omega_{0}/\omega_{b}=0.05$, and $\omega_{b}T=10000$. (b) One period of the equal-time second-order correlation functions $g_{N}^{(2)}(t,t)$ (zoomed-in plot of shaded area). $t_{s1}$ and $t_{s2}$ correspond to the maximum value of $g_{1}^{(2)}(t,t)$ and the minimum value of $g_{2}^{(2)}(t,t)$, respectively. (c) The time-delay second-order correlation functions $g_{N}^{(2)}(t_{sN},t_{sN}+\tau)$ [$t_{s1}$ and $t_{s2}$ are indicated in Fig.~\ref{Fig5}(b)] with $N=1$ (red) and $N=2$ (blue). (d) One period of the equal-time second-order correlation functions $g_{N}^{(2)}(t,t)$ with $N=1$ (red) and $N=3$ (green) when $\lambda/\omega_{b}\approx0.645$, $\Omega_{0}/\omega_{b}=0.1$, and $\omega_{b}T=12000$. (e) The time-delay second-order correlation functions $g_{N}^{(2)}(t_{qN},t_{qN}+\tau)$ [$t_{q1}$ and $t_{q3}$ are indicated in Fig.~\ref{Fig5}(d)] with $N=1$ (red) and $N=3$ (green). Other common used parameters are the same as those in Fig.~\ref{Fig3}.}
\label{Fig5}
\end{figure}
%%%%%%%%%%%%%%%%%%%%%%%

To characterize the statistical properties of the dynamical $N$-photon bundle emission, we consider the generalized $n$th-order correlation functions defined by~\cite{munoz2014Emitters}:
\begin{equation}\label{corrfun}
g_{N}^{(n)}(t_{1},\ldots,t_{n})=\frac{\langle\mathcal{T_{-}}\{\prod_{i=1}^{n}b^{\dagger N}(t_{i})\}\mathcal{T_{+}}\{\prod_{i=1}^{n}b^{N}(t_{i})\}\rangle}{\prod_{i=1}^{n}\langle b^{\dagger N}b^{N}\rangle(t_{i})},
\end{equation}
where $\mathcal{T_{\pm}}$ represents the time-ordering operators and we neglect the time interval within photon bundle. Note that the standard correlation function $g^{(n)}(t_{1},\ldots,t_{n})$ corresponds to the case of $N=1$ in the generalized correlation function (\ref{corrfun}). In Fig.~\ref{Fig5}(a), we show the evolution of the equal-time second-order correlation functions $g_{1}^{(2)}(t,t)$ and $g_{2}^{(2)}(t,t)$ at $\lambda/\omega_{b}\approx0.765$. We can see that the correlation functions periodically change under the action of two Gaussian-pulse sequences. Figure~\ref{Fig5}(b) shows one period of the correlation functions $g_{1}^{(2)}(t,t)$ and $g_{2}^{(2)}(t,t)$. It can be found that $g_{1}^{(2)}(t,t)$ reaches maximum at time $t=t_{s1}$ and $g_{1}^{(2)}(t_{s1},t_{s1})>1$, which indicates the super-Poisson distribution of a single photon. Moreover, $g_{2}^{(2)}(t,t)$ reaches its minimum at time $t=t_{s2}$, and $g_{2}^{(2)}(t_{s2},t_{s2})<1$ indicating the sub-Poisson distribution of two-photon bundle. To further describe the statistical properties of the dynamical two-photon bundle emission, we show in Fig.~\ref{Fig5}(c) the time-delay second-order correlation functions $g_{1}^{(2)}(t_{s1},t_{s1}+\tau)$ and $g_{2}^{(2)}(t_{s2},t_{s2}+\tau)$ with delayed time $\tau$. It can be seen that $g_{1}^{(2)}(t_{s1},t_{s1})>g_{1}^{(2)}(t_{s1},t_{s1}+\tau)$ and $g_{2}^{(2)}(t_{s2},t_{s2})<g_{2}^{(2)}(t_{s2},t_{s2}+\tau)$, which shows bunching of a single photon, but antibunching between two-photon bundle. In addition, we also study the statistical properties of the dynamical three-photon bundle emission in Figs.~\ref{Fig5}(d) and~\ref{Fig5}(e) when $\lambda/\omega_{b}\approx0.645$. It can be seen from Fig.~\ref{Fig5}(d) that $g_{1}^{(2)}(t_{q1},t_{q1})>1$ [$g_{3}^{(2)}(t_{q3},t_{q3})<1$], which indicates the super-Poisson (sub-Poisson) distribution of a single photon (three-photon bundle). In Fig.~\ref{Fig5}(e), we observe that $g_{1}^{(2)}(t_{q1},t_{q1})>g_{1}^{(2)}(t_{q1},t_{q1}+\tau)$ and $g_{3}^{(2)}(t_{q3},t_{q3})<g_{3}^{(2)}(t_{q3},t_{q3}+\tau)$. This means bunching of a single photon and antibunching between three-photon bundle.

It should be pointed out that the photon emission in this system occurs due to the intrinsic decay of the resonator, which means that the time of each photon emission has an inherent uncertainty~\cite{Lu1989Effects}. However, in this work we consider the case where the time interval between any two consecutive STIRAP pulses is much larger than the time duration of $N$-photon emission in a bundle. Therefore, the time interval within the $N$ photons can be ignored on the timescales of two Gaussian-pulse sequences. In this sense, the $N$ photons can be approximately considered as a photon bundle. This point can also be confirmed from the correlation function~(\ref{corrfun}), which indicates that the $N$ photons are emitted simultaneously.

\section{Discussion and Conclusion}
Finally, we present some discussions on the experimental implementation of the present physical scheme. In this work, we have considered a circuit-QED system with a qubit longitudinally coupled to a microwave resonator~\cite{Romero2012Ultrafast,Didier2015Fast,Billangeon2015Circuit,Billangeon2015Scalable,Richer2016Circuit,Richer2017Inductively,Touzard2019Gated,Ikonen2019Qubit,
Hacohen2016Quantum,Eddins2018Stroboscopic}. Meanwhile, we should emphasize that the physics mechanism in this work is general, and hence it can be implemented with other physical platforms, with which the same interactions and drivings can be implemented. For example, our proposal can be implemented in the semiconductor system with a quantum dot coupled to a acoustic nanocavity~\cite{bin2020Phonon}. Below, we present some analyses on the suggested experimental parameters. It can be seen from Eq.~(\ref{miniequ}) that the ultrastrong longitude coupling strength is the key to implement the present scheme. In our simulations, we used the following parameters: $\lambda/\omega_{b}\approx0.765$ or $0.645$, $\Omega_{0}/\omega_{b}=0.05\sim0.1$, $\kappa/\omega_{b}=0.0004\sim0.0008$, and $\gamma/\omega_{b}=0.002$ [e.g., $\omega_{b}=2\pi\times5\,\textrm{GHz}$, $\lambda\approx2\pi\times(3.825$ or $3.225)\,\textrm{GHz}$, $\Omega_{0}=2\pi\times(0.25\sim0.5)\,\textrm{GHz}$, $\kappa=2\pi\times(2\sim4)\,\textrm{MHz}$, and $\gamma=2\pi\times10\,\textrm{MHz}$]. These suggested parameters are experimentally accessible in current circuit-QED systems~\cite{Touzard2019Gated,Ikonen2019Qubit,Hacohen2016Quantum,Eddins2018Stroboscopic}. In particular, the experimental realization of the ultrastrong even deep-strong coupling regimes of circuit-QED systems has been reported~\cite{Niemczyk2010Circuit,Forn2010Observation,Yoshihara2017Superconducting,Forn2017Ultrastrong,Chen2017Single}.

In conclusion, we have proposed the dynamical $N$-photon bundle emission and designed a feasible scheme to implement the physical process in a circuit-QED system based on the feature of the STIRAP. We have demonstrated that the dynamical two-photon (three-photon) bundle emission behaves as the bunching of a single photon and antibunching between two-photon (three-photon) bundle when the time interval between the consecutive pulses is much larger than the photon lifetime $T\gg1/\kappa$. Our work opens up a route to achieve multiphoton source emitter, which are very useful for quantum information processing and medical applications.

\section*{Acknowledgments}
J.-Q.L. was supported in part by National Natural Science Foundation of China (Grants No.~12175061, No.~12247105, and No.~11935006) and the Science and Technology Innovation Program of Hunan Province (Grants No.~2021RC4029 and No.~2020RC4047). Y.L. was supported in part by National Natural Science Foundation of China (Grants No.~12074030 and No.~12274107). F.Z. was supported in part by the National Natural Science Foundation of China (Grants No.~12147109 and No.~U2230402) and the China Postdoctoral Science Foundation (Grant No.~2021M700360).

\appendix
\setcounter{section}{1}
\section*{Appendix: Derivation of the dressed-state master equation~(\ref{SME})}
In this Appendix, we derive the dressed-state master equation given in Eq.~(\ref{SME}), which governs the evolution of the circuit-QED system in the ultrastrong-coupling regime~\cite{Diaz2019Ultrastrong,Kockum2019Ultrastrong,Boite2020Theoretical}. We consider the case where the resonator and the qubit are coupled with two independent heat baths, then the Hamiltonian of the whole system including the circuit-QED system and its baths reads
\begin{equation}
H^{\prime}=H_{sd}+H_{b}^{B}+H_{\sigma}^{B}+H_{bB}+H_{\sigma B},
\end{equation}
where $H_{sd}=H_{s}+H_{d}$ and the Hamiltonian related to the baths are given by
\begin{eqnarray}
H_{b}^{B} & =&\sum_{j}\omega_{j}b_{j}^{\dagger}b_{j},\quad
H_{bB} =\left(b^{\dagger}+b\right)\sum_{j}\lambda_{j}^{b}(b_{j}^{\dagger}+b_{j}),\\
H_{\sigma}^{B} & =&\sum_{k}\omega_{k}c_{k}^{\dagger}c_{k},\quad
H_{\sigma B} =\sigma_{+}\sum_{k}\lambda_{k}^{\sigma}c_{k}+\sigma_{-}\sum_{k}\lambda_{k}^{\sigma}c_{k}^{\dagger}.
\end{eqnarray}
Here the creation and annihilation operators $b_{j}^{\dagger}$ ($c_{k}^{\dagger}$) and $b_{j}$ ($c_{k}$) describe the $j$th ($k$th) mode with resonance frequency $\omega_{j}$ ($\omega_{k}$) of the resonator (qubit) bath, and $\lambda_{j}^{b}$ ($\lambda_{k}^{\sigma}$) is the coupling strength between the resonator (qubit) and the $j$th ($k$th) mode of the corresponding heat bath. In the interaction picture with respect to $H_{0}^{\prime}=H_{sd}+H_{b}^{B}+H_{\sigma}^{B}$, the equation of motion for the density operator $\rho_{SB}(t)$ of the whole system is given by $\dot{\rho}_{SB}(t)=i\left[\rho_{SB}(t),H_{I}(t)\right]$, where the interaction Hamiltonian $H_{I}(t)$ is given by
\begin{eqnarray}\label{SHI}
H_{I}(t) & =&e^{iH_{0}^{\prime}t}(H_{bB}+H_{\sigma B})e^{-iH_{0}^{\prime}t}\nonumber \\
&=&[\tilde{b}^{\dagger}(t)+\tilde{b}(t)][\tilde{\Gamma}_{b}^{\dagger}(t)+\tilde{\Gamma}_{b}(t)]
+[\tilde{\sigma}_{+}(t)\tilde{\Gamma}_{\sigma}(t)+\tilde{\sigma}_{-}(t)\tilde{\Gamma}_{\sigma}^{\dagger}(t)].
\end{eqnarray}
In Eq.~(\ref{SHI}), we introduce the operators
\begin{eqnarray}\label{Soper}
\tilde{b}(t) &=&e^{iH_{sd}t}be^{-iH_{sd}t}\approx e^{iH_{s}t}be^{-iH_{s}t}, \quad
\tilde{\Gamma}_{b}(t) =\sum_{j}\lambda_{j}^{b}b_{j}e^{-i\omega_{j}t},\nonumber \\
\tilde{\sigma}_{-}(t) &=& e^{iH_{sd}t}\sigma_{-}e^{-iH_{sd}t}\approx e^{iH_{s}t}\sigma_{-}e^{-iH_{s}t}, \quad
\tilde{\Gamma}_{\sigma}(t) =\sum_{k}\lambda_{k}^{\sigma}c_{k}e^{-i\omega_{k}t}.
\end{eqnarray}
Note that in the derivation of Eq.~(\ref{Soper}), we have ignored the driving term $H_{d}$ when $\Omega_{0}\ll\{\omega_{0},\omega_{b}$\}~\cite{Ridolfo2012Photon,bin2021ParitySymmetryProtected}, namely, we take $H_{sd}\approx H_{s}=\omega_{b}b^{\dagger}b+\omega_{0}\sigma_{+}\sigma_{-}+\lambda\sigma_{+}\sigma_{-}(b^{\dagger}+b)$ in the derivation of the quantum master equation.

Under the Born-Markov approximation~\cite{scully1997Quantum}, the quantum master equation of the reduced density matrix $\rho_{I}(t)=\mathrm{Tr}_{B}[\rho_{SB}(t)]$ in the interaction picture can be derived as
\begin{equation}\label{rhoIS}
\dot{\rho}_{I}(t)=-\int_{0}^{\infty}ds\mathrm{Tr}_{B}([H_{I}(t),[H_{I}(t-s),\rho_{I}(t)\otimes\rho_{B}]]),
\end{equation}
where $\mathrm{Tr}_{B}$ denotes the trace operation over the bath mode and $\rho_{B}$ is the initial density operator of the bath. By substituting $H_{I}(t)$ and $H_{I}(t-s)$ into Eq.~(\ref{rhoIS}) and applying the rotating-wave approximation to neglect the fast-oscillating terms, the quantum master equation of the reduced density matrix $\rho_{I}(t)$ in the interaction picture can be obtained as
\begin{equation}
\dot{\rho}_{I}(t)=\mathcal{L}_{b}[\rho_{I}(t)]+\mathcal{L}_{\sigma}[\rho_{I}(t)],
\end{equation}
where the two dissipation parts are given by
\begin{eqnarray}\label{Srhob}
\mathcal{L}_{b}[\rho_{I}(t)]&=& -\int_{0}^{\infty}ds[\tilde{b}^{\dagger}(t)+\tilde{b}(t)][\tilde{b}^{\dagger}(t-s)+\tilde{b}(t-s)]\rho_{I}(t)Y(s)+\nonumber \\
&& \int_{0}^{\infty}ds[\tilde{b}^{\dagger}(t)+\tilde{b}(t)]\rho_{I}(t)[\tilde{b}^{\dagger}(t-s)+\tilde{b}(t-s)]R(s)+\mathrm{H.c.},
\end{eqnarray}
and
\begin{eqnarray}\label{Srhosigma}
\mathcal{L}_{\sigma}[\rho_{I}(t)]&=& -\int_{0}^{\infty}ds\tilde{\sigma}_{+}(t)\tilde{\sigma}_{-}(t-s)\rho_{I}(t)Y_{-}(s)\nonumber \\
&&
-\int_{0}^{\infty}ds\tilde{\sigma}_{-}(t)\tilde{\sigma}_{+}(t-s)\rho_{I}(t)Y_{+}(s)\nonumber \\
&& +\int_{0}^{\infty}ds\tilde{\sigma}_{+}(t)\rho_{I}(t)\tilde{\sigma}_{-}(t-s)R_{-}(s)\nonumber \\
&&
+\int_{0}^{\infty}ds\tilde{\sigma}_{-}(t)\rho_{I}(t)\tilde{\sigma}_{+}(t-s)R_{+}(s)+\mathrm{H.c.}.
\end{eqnarray}
In Eqs.~(\ref{Srhob}) and~(\ref{Srhosigma}), the correlation functions of the resonator and qubit baths are, respectively, defined as
\begin{eqnarray}
Y(s)&=& \mathrm{Tr}_{B}
[\tilde{\Gamma}_{b}^{\dagger}(t)\tilde{\Gamma}_{b}(t-s)\rho_{B}+\tilde{\Gamma}_{b}(t)\tilde{\Gamma}_{b}^{\dagger}(t-s)\rho_{B}] \nonumber\\
&=&\sum_{j}\vert\lambda_{j}^{b}\vert^{2}e^{i\omega_{j}s}\bar{n}(\omega_{j},T_{b})+\sum_{j}\vert\lambda_{j}^{b}\vert^{2}
e^{-i\omega_{j}s}[\bar{n}(\omega_{j},T_{b})+1],\nonumber\\
R(s)&=&\mathrm{Tr}_{B}
[\tilde{\Gamma}_{b}^{\dagger}(t)\rho_{B}\tilde{\Gamma}_{b}(t-s)+\tilde{\Gamma}_{b}(t)\rho_{B}\tilde{\Gamma}_{b}^{\dagger}(t-s)] \nonumber\\
&=&\sum_{j}\vert\lambda_{j}^{b}\vert^{2}e^{i\omega_{j}s}[\bar{n}(\omega_{j},T_{b})+1]+\sum_{j}\vert\lambda_{j}^{b}\vert^{2}
e^{-i\omega_{j}s}\bar{n}(\omega_{j},T_{b}),
\end{eqnarray}
and
\begin{eqnarray}
Y_{-}(s)&=&\mathrm{Tr}_{B}[\tilde{\Gamma}_{\sigma}(t)\tilde{\Gamma}_{\sigma}^{\dagger}(t-s)\rho_{B}]
=\sum_{k}\vert\lambda_{k}^{\sigma}\vert^{2}e^{-i\omega_{k}s}[\bar{n}(\omega_{k},T_{\sigma})+1],\nonumber\\
Y_{+}(s)&=&\mathrm{Tr}_{B}[\tilde{\Gamma}_{\sigma}^{\dagger}(t)\tilde{\Gamma}_{\sigma}(t-s)\rho_{B}]
=\sum_{k}\vert\lambda_{k}^{\sigma}\vert^{2}e^{i\omega_{k}s}\bar{n}(\omega_{k},T_{\sigma}),\nonumber\\
R_{-}(s)&=&\mathrm{Tr}_{B}[\tilde{\Gamma}_{\sigma}(t)\rho_{B}\tilde{\Gamma}_{\sigma}^{\dagger}(t-s)]
=\sum_{k}\vert\lambda_{k}^{\sigma}\vert^{2}e^{-i\omega_{k}s}\bar{n}(\omega_{k},T_{\sigma}),\nonumber\\
R_{+}(s)&=&\mathrm{Tr}_{B}[\tilde{\Gamma}_{\sigma}^{\dagger}(t)\rho_{B}\tilde{\Gamma}_{\sigma}(t-s)]
=\sum_{k}\vert\lambda_{k}^{\sigma}\vert^{2}e^{i\omega_{k}s}[\bar{n}(\omega_{k},T_{\sigma})+1],
\end{eqnarray}
where $\bar{n}(\omega_{j},T_{b})=1/[\exp(\hbar\omega_{j}/k_{B}T_{b})-1]$ and $\bar{n}(\omega_{k},T_{\sigma})=1/[\exp(\hbar\omega_{k}/k_{B}T_{\sigma})-1]$ are, respectively, the average occupation number associated with the resonator and the qubit. Below we will derive the contributions from the resonator and qubit baths in detail.

\textit{Resonator bath contribution---}For the Hamiltonian $H_{s}=\omega_{b}b^{\dagger}b+\omega_{0}\sigma_{+}\sigma_{-}+\lambda\sigma_{+}\sigma_{-}(b^{\dagger}+b)$, its eigenvalues are given by $\varepsilon_{g,n}=n\omega_{b}$ and $\varepsilon_{e,n}=n\omega_{b}+\omega_{0}-\lambda^{2}/\omega_{b}$, with the corresponding eigenstates $\vert g,n\rangle$ and $\vert e,\tilde{n}\rangle$. By using the eigenstates of the Hamiltonian $H_{s}$, the time-dependent operator $\tilde{b}(t)$ can be expressed as
\begin{eqnarray}
\tilde{b}(t)
&=&\sum_{m=0}^{\infty}e^{-i\omega_{b}t}b\vert g,m\rangle\langle g,m\vert+\sum_{m=0}^{\infty}e^{-i\omega_{b}t}(b+\lambda/\omega_{b})\vert e,\tilde{m}\rangle \langle e,\tilde{m}\vert\nonumber\\
&&-(\lambda/\omega_{b})\sum_{m=0}^{\infty}\vert e,\tilde{m}\rangle\langle e,\tilde{m}\vert,
\end{eqnarray}
where we use the relation $b\vert\tilde{m}\rangle =\sqrt{m}\vert\widetilde{m-1}\rangle -(\lambda/\omega_{b})\vert\tilde{m}\rangle$.
Therefore, we obtain the relations
\begin{equation}
\tilde{b}(t) =e^{-i\omega_{b}t}[b+(\lambda/\omega_{b})\sigma_{+}\sigma_{-}]-(\lambda/\omega_{b})\sigma_{+}\sigma_{-}.
\end{equation}

Assume that the spectral density of the resonator bath is Ohmic, i.e., $J_{b}(\omega)=\sum_{j}\vert\lambda_{j}^{b}\vert^{2}\delta(\omega-\omega_{j})$. It appears as $J_{b}(\omega)=\kappa\omega/(2\pi\omega_{b})$ in the continuum limit of bath frequency, where $\kappa=2\pi J_{b}(\omega_{b})$ is the decay rate of the resonator. Using the formula~\cite{Hu2015Quantum}
\begin{eqnarray}
&&\int_{0}^{\infty}dse^{i\omega_{b}s}Y(s)=\frac{\kappa}{2}[\bar{n}(\omega_{b},T_{b})+1],\quad
\int_{0}^{\infty}dsY(s)=\frac{\kappa}{2}\frac{k_{B}T_{b}}{\omega_{b}},\nonumber\\
&&\int_{0}^{\infty}dse^{-i\omega_{b}s}Y(s)=\frac{\kappa}{2}\bar{n}(\omega_{b},T_{b}),\quad
\int_{0}^{\infty}dse^{i\omega_{b}s}R(s)=\frac{\kappa}{2}\bar{n}(\omega_{b},T_{b}),\nonumber\\
&&\int_{0}^{\infty}dse^{-i\omega_{b}s}R(s)=\frac{\kappa}{2}[\bar{n}(\omega_{b},T_{b})+1],\quad
\int_{0}^{\infty}dsR(s)=\frac{\kappa}{2}\frac{k_{B}T_{b}}{\omega_{b}},
\end{eqnarray}
and omitting the fast-oscillating terms, the contribution of the resonator bath can then be obtained as
\begin{eqnarray}\label{Lb}
\mathcal{L}_{b}[\rho_{I}(t)]&=& \kappa[\bar{n}(\omega_{b},T_{b})+1]\mathcal{L}[b+(\lambda/\omega_{b})\sigma_{+}\sigma_{-}]\rho_{I}(t)\nonumber\\
&&+\kappa\bar{n}(\omega_{b},T_{b})\mathcal{L}[b^{\dagger}+(\lambda/\omega_{b})\sigma_{+}\sigma_{-}]\rho_{I}(t)\nonumber\\
&&+4\kappa(k_{B}T_{b}/\omega_{b})(\lambda/\omega_{b})^{2}\mathcal{L}[\sigma_{+}\sigma_{-}]\rho_{I}(t),
\end{eqnarray}
where $\mathcal{L}[o]\rho_{I}(t)=[2o\rho_{I}(t)o^{\dagger}-o^{\dagger}o\rho_{I}(t)-\rho_{I}(t)o^{\dagger}o]/2$ is the Lindblad superoperator for operator $o$ in the interaction picture.

\textit{Qubit bath contribution---}Similarly, by using the eigenstates of the Hamiltonian $H_{s}$, the time-dependent operator $\tilde{\sigma}_{-}(t)$ can be expressed as
\begin{eqnarray}
\tilde{\sigma}_{-}(t)&=&\sum_{n,m=0}^{\infty}e^{i(\varepsilon_{g,n}-\varepsilon_{e,m})t}\langle n\vert\tilde{m}\rangle\vert g,n\rangle\langle e,\tilde{m}\vert=\sum_{n,m=0}^{\infty}e^{-i\Delta_{n,m}t}A_{n,m},
\end{eqnarray}
where we define $\Delta_{n,m}=\varepsilon_{e,m}-\varepsilon_{g,n}=(m-n)\omega_{b}+\omega_{0}-\lambda^{2}/\omega_{b}$ and $A_{n,m}=\langle n\vert\tilde{m}\rangle\vert g,n\rangle\langle e,\tilde{m}\vert$. The contribution of the qubit bath can be further expressed as
\begin{eqnarray}
\mathcal{L}_{\sigma}[\rho_{I}(t)]
&=& -\int_{0}^{\infty}ds\sum_{n,m=0}^{\infty}e^{i\Delta_{n,m}t}A^{\dagger}_{n,m}\sum_{p,q}e^{-i\Delta_{p,q}(t-s)}A_{p,q}\rho_{I}(t)Y_{-}(s)\nonumber\\
&&-\int_{0}^{\infty}ds\sum_{n,m=0}^{\infty}e^{-i\Delta_{n,m}t}A_{n,m}\sum_{p,q}e^{i\Delta_{p,q}(t-s)}A^{\dagger}_{p,q}\rho_{I}(t)Y_{+}(s)\nonumber \\
&& +\int_{0}^{\infty}ds\sum_{n,m=0}^{\infty}e^{i\Delta_{n,m}t}A^{\dagger}_{n,m}\rho_{I}(t)\sum_{p,q}e^{-i\Delta_{p,q}(t-s)}A_{p,q}R_{-}(s)\nonumber\\
&&+\int_{0}^{\infty}ds\sum_{n,m=0}^{\infty}e^{-i\Delta_{n,m}t}A_{n,m}\rho_{I}(t)\sum_{p,q}e^{i\Delta_{p,q}(t-s)}A^{\dagger}_{p,q}R_{+}(s)\nonumber\\
&&+\mathrm{H.c.}.
\end{eqnarray}

Under the condition of $\omega_{0}\gg\omega_{b}$, we further assume that the spectral density of the qubit bath, defined as $J_{0}(\omega)=\sum_{k}\vert\lambda_{k}^{\sigma}\vert^{2}\delta(\omega-\omega_{k})$, is a slow varying function in the vicinity of $\omega=\omega_{0}$. Then the spectral density can be approximated as $J_{0}(\omega)\equiv\gamma/2\pi$ in the entire range of the photon sidebands. In this case, we can obtain the following relations~\cite{Hu2015Quantum}:
\begin{eqnarray}\label{SQDbath}
\int_{0}^{\infty}dse^{i\Delta_{p,q}s}Y_{-}(s)&\approx&\frac{\gamma}{2}[\bar{n}(\omega_{0},T_{\sigma})+1],\nonumber\\
\int_{0}^{\infty}dse^{-i\Delta_{p,q}s}Y_{+}(s)&\approx&\frac{\gamma}{2}\bar{n}(\omega_{0},T_{\sigma}),\nonumber\\
\int_{0}^{\infty}dse^{-i\Delta_{p,q}s}R_{+}(s)&\approx&\frac{\gamma}{2}[\bar{n}(\omega_{0},T_{\sigma})+1],\nonumber\\
\int_{0}^{\infty}dse^{i\Delta_{p,q}s}R_{-}(s)&\approx&\frac{\gamma}{2}\bar{n}(\omega_{0},T_{\sigma}).
\end{eqnarray}
By using Eq.~(\ref{SQDbath}) and neglecting the fast-oscillating terms, the contribution of the qubit bath can be obtained as
\begin{equation}\label{Lsigma}
\mathcal{L}_{\sigma}[\rho_{I}(t)]
= \gamma\bar{n}(\omega_{0},T_{\sigma})\mathcal{L}[\tilde{\sigma}_{+}(t)]\rho_{I}(t)+\gamma[\bar{n}(\omega_{0},T_{\sigma})+1]\mathcal{L}
[\tilde{\sigma}_{-}(t)]\rho_{I}(t).
\end{equation}

Hence, the dressed-state master equation for the reduced density matrix $\rho_{I}(t)$ in the interaction picture is given by
\begin{equation}
\dot{\rho}_{I}(t)=\mathcal{L}_{b}[\rho_{I}(t)]+\mathcal{L}_{\sigma}[\rho_{I}(t)],
\end{equation}
where $\mathcal{L}_{b}[\rho_{I}(t)]$ and $\mathcal{L}_{\sigma}[\rho_{I}(t)]$ are given in Eqs.~(\ref{Lb}) and~(\ref{Lsigma}). In the zero-temperature case and transforming back to the Schr\"{o}dinger picture [$\rho_{s}(t)=e^{-iH_{sd}t}\rho_{I}(t)e^{iH_{sd}t}$], we obtain the dressed-state master equation of the circuit-QED system as
\begin{equation}
\frac{d\rho_{s}(t)}{dt}=i[\rho_{s}(t),H_{sd}]+\kappa\mathcal{L}[b+(\lambda/\omega_{b})\sigma_{+}\sigma_{-}]\rho_{s}(t)+\gamma\mathcal{L}[\sigma_{-}]\rho_{s}(t).
\end{equation}
In the rotating frame with respect to $\omega_{0}\sigma_{+}\sigma_{-}$, the dressed-state master equation for the density operator
$\rho$ in the ultrastrong-coupling regime can be expressed as
\begin{equation}
\frac{d\rho(t)}{dt}=i[\rho(t),H]+\kappa\mathcal{L}[b+(\lambda/\omega_{b})\sigma_{+}\sigma_{-}]\rho(t)+\gamma\mathcal{L}[\sigma_{-}]\rho(t),
\end{equation}
where $H$ is given by Eq.~(\ref{H}).

\section*{Data availability statement}
All data that support the findings of this study are included within the article (and any supplementary files).

%\bibliography{RefNPBE}

\end{document}